\documentclass{aa}  
\usepackage{graphicx}
\usepackage{subfigure}
\usepackage{amsmath}
\usepackage{natbib}
\usepackage{bm}
\usepackage{epstopdf}
\usepackage{txfonts}
\usepackage{natbib}
\usepackage{longtable}
\bibpunct{(}{)}{;}{a}{}{,} 

\def\mso{\,\mathrm{M}_\odot}
\def\rso{\,{\rm R}_\odot}
\def\lso{\,{\rm L}_\odot}

\def\kms{\, {\rm km}\, {\rm s}^{-1}}

\def\simle{\mathrel{\hbox{\rlap{\hbox{\lower4pt\hbox{$\sim$}}}\hbox{$<$}}}}
\def\simgr{\mathrel{\hbox{\rlap{\hbox{\lower4pt\hbox{$\sim$}}}\hbox{$>$}}}}

\def\grad{\nabla}
\def\adrad{\nabla_{\mathrm{\!rad}}}

\def\hp{H_{\mathrm{P}}}

\def\c2{^{12}{\rm C}}
\def\c3{^{13}{\rm C}}
\def\n14{^{14}{\rm N}}
\def\c1213{^{12}{\rm C}/^{13}{\rm C}}
\def\he3he4{^3{\rm He}/^4{\rm He}}

\def\vca{\varv_{\mathrm {c}}}

\def\vin{\varv_{\,\mathrm{\infty}}}

\def\tauspot{$t_{\mathrm{spot}}$}
\def\ttwenty{6.41$\times10^6$ }
\def\tsixty{2.37$\times10^6$ }
\newcommand{\EQ}{\begin{equation}}
\newcommand{\EE}{\end{equation}}
\newcommand{\EQA}{\begin{eqnarray}}
\newcommand{\EEA}{\end{eqnarray}}

\newcommand\T{\rule{0pt}{2.0ex}}       
\def\ttwenty{6.41$\times10^6$ }
\def\tsixty{2.37$\times10^6$ }
\begin{document}
  \title{Magnetic spots on hot massive stars} 
  \author{M. Cantiello
	  \&
	  J. Braithwaite}
  \offprints{M. Cantiello}

  \institute{Argelander-Institut f\"ur Astronomie der Universit\"at Bonn, Auf dem H\"ugel 71, D--53121 Bonn, Germany\\
	      \email{cantiello@astro.uni-bonn.de}    
	    }

  \date{ }

  \abstract
  {Hot luminous stars show a variety of phenomena in their photospheres and winds
which still lack clear physical explanation. Among these phenomena are
photospheric turbulence, line profile variability (LPV), non-thermal emission, non-radial pulsations, 
discrete absorption components (DACs) and wind
clumping.  Cantiello et al. (2009) argued that a convection zone close to the stellar surface could be responsible for some of these phenomena.  This convective zone is caused by a peak in the opacity associated with iron-group elements and is referred to as the ``iron convection zone'' (FeCZ). } 
  {Assuming dynamo action producing magnetic fields at equipartition in the FeCZ, we investigate the occurrence of subsurface magnetism in OB stars. Then we study the surface emergence of these magnetic fields and    discuss possible observational signatures of magnetic spots.} 
  {Simple estimates are made using the subsurface properties  of massive stars, as calculated in 1D stellar evolution models.}
  {We find that magnetic fields of sufficient amplitude to affect the wind could emerge at the surface via magnetic buoyancy. 
  While at this stage it is difficult to predict the geometry of these features, we show that 
  magnetic spots of size comparable to the local pressure scale height can manifest themselves as hot, bright spots. } 
 {Localized magnetic fields could be widespread in those early type stars that have subsurface convection.  This type of surface magnetism could be responsible for  
  photometric variability and play a role in X-ray emission and wind clumping.}
  
  \keywords{}
  \maketitle


\section{Introduction}
There is growing direct and indirect evidence that magnetic fields are present at the surface of massive stars. Direct measurements have been obtained using the Zeeman effect for O and early B stars 
 \citep[see e.g.][for a review]{2010arXiv1010.2248P}.  As the number of known  magnetic massive stars is small,  these observations have not yet been able to answer the question of whether magnetism in these stars is fundamentally different from that in  the intermediate-mass stars, i.e. late B and A stars (the so-called Ap and Bp stars; see e.g. \citet{2009ASPC..405..473M} for a review, or \citet{2005LNP...664..183M} for a broader review of stellar magnetism), or just a continuation of the same phenomenon. Ap and Bp stars -- accounting for some percentage ($\sim5-10\%$) of main-sequence A and late B stars -- display a magnetic field which is static, in the rotating frame (these stars appear to rotate as solid bodies), large-scale, and with strengths of $200$G to $100$kG. These fields are thought to be `fossil' remnants left over from formation: magnetohydrostatic equilibria which survive for the entire stellar lifetime \citep{1945MNRAS.105..166C,2006A&A...450.1077B,2010ApJ...724L..34D}. However, since direct detection methods are biased toward the discovery of large scale magnetic fields, small-scale fields generated by a contemporary dynamo may be present in the rest of the population. Whilst in A stars even small-scale fields would have to be rather weak  to escape detection \citep{2010A&A...523A..41P}, the detection limits on small-scale fields in more massive stars are much higher \citep[e.g.][]{2008A&A...483..857S}.

In addition there is a variety of indirect evidence for such magnetic fields at the surface of OB stars, coming from observations of surface and wind phenomena. For example, discrete absorption components (DACs) and line profile variability (LPV) have been often associated with the presence of surface magnetic fields, even if non radial pulsation (NRP) is  also a possibility \citep[e.g.,][]{1996ApJ...462..469C,1996ApJS..103..475F,1997A&A...327..281K,2005ASPC..337..114H}. Interestingly these phenomena seem to be ubiquitous among early type stars \citep[e.g.,][]{1989ApJS...69..527H,1996A&AS..116..257K,1996ApJS..103..475F,2002A&A...388..587P}. Therefore one may wonder if magnetic fields in massive stars are more widespread than previously thought. 
 It is important to stress that small scale magnetic fields of amplitude $\sim$10...100 G are in principle able to affect the wind of OB stars \citep{2002ApJ...576..413U}.
Moreover, theoretical considerations point to a difference between massive and intermediate-mass stars. In the latter, a dynamo would have to operate either in the radiative envelope, powered by differential rotation \citep{2002A&A...381..923S} which itself would have to be 
sustained by some physical process;
or in the convective core, in which case the resulting field would have to make its way upwards through the radiative envelope \citep{1989MNRAS.236..629M,2003ApJ...586..480M} which seems to take place only on an embarrassingly long timescale. In contrast to this, recently \citet[][hereafter C09]{2009A&A...499..279C} studied the properties of
convective regions in the envelopes of hot, massive stars, caused by an opacity peak associated with iron \citep{1992ApJ...397..717I}. These convective zones (hereafter FeCZ), which are not present in intermediate-mass stars, could host a dynamo, producing magnetic fields visible at the surface. 

The reality and importance of subsurface convective layers in early-type stars has been recently stressed by observations of stellar surface properties. 
The idea that  the observed microturbulence is caused by the presence of the FeCZ (C09) is supported by recent observations  \citep[][Przybilla, priv.comm.]{2010MNRAS.404.1306F}.
In the Sun stochastic convective motions excite non-radial pulsations \citep{1970ApJ...162..993U,1971ApL.....7..191L}. In analogy with the Sun, it has been suggested that solar-like oscillations are excited in hot, massive stars by the FeCZ \citep[C09,][]{2010A&A...510A...6B}.
 Intriguingly, non-radial pulsations that are compatible with stochastic excitation from subsurface convection
have recently been found by COROT in both a B \citep{2009Sci...324.1540B} and
an O star \citep{2010A&A...519A..38D}, but see also \citet{2011MNRAS.tmp..298B}.

C09 pointed out that the FeCZ could host dynamo action very close to the
surface. Simulations of turbulent convection in the presence of shear and
rotation show that dynamo action is possible and can produce magnetic fields
reaching equipartition on scales larger than that of the convection \citep[e.g.,][]{2008A&A...491..353K,2010arXiv1009.4462C,2011arXiv1102.3598G}. In this paper we
consider the emergence of magnetic fields produced in a subsurface convective
layer. In the following section we briefly review the properties of the FeCZ,
before examining  in Sect.~\ref{spots} the magnetic fields produced within them and mechanisms for
those fields to reach the surface. In Sect.~\ref{consequences} we look at observable effects on the surface, and finally we discuss the results and conclude in Sect.~\ref{disc} and \ref{conc}. 

\begin{table*}

\begin{minipage}[t]{\textwidth}
\centering
\caption{Properties of the outer layers in a 20$\mso$ and 60$\mso$ model main-sequence stars of solar metallicity at ages \ttwenty yr and \tsixty yr respectively, corresponding to about 90\% and 80\% of the main-sequence lifetime.}
\label{table}
\renewcommand{\footnoterule}{}

\begin{tabular}{l c | c c c c c c  c c c c c c}
\hline\hline
   $M_{\rm{ini}}$    & $R_{\star}$    \T  & $R_{\mathrm{FeCZ}}$\footnote{Radial coordinate of the top of the FeCZ.} & $\Delta R_{\mathrm{FeCZ}}$\footnote{Radial extension of the FeCZ.}    &      $\hp$\footnote{Pressure scale height at top/bottom of the FeCZ.} & $\vca$\footnote{Maximum of the convective velocity inside the FeCZ.}  & $\rho$\footnote{Density at $\vca$.} & $\Delta M_{\rm FeCZ}$\footnote{Mass contained in the convective region.}  & $\Delta M_{\rm top}$\footnote{Mass in the radiative layer between the stellar surface and the upper boundary of the convective zone.}    & $\tau_{\rm turn}$\footnote{Convective turnover time, $\tau_{\rm turn}:= \hp/\vca$.} &   $\tau_{\rm conv}$\footnote{Time that a piece of stellar material spends 
inside a convective region, $\tau_{\rm conv}:= \Delta M_{\rm FeCZ}/\dot M$.} & $ \dot{M}$ \\
 $\mso$ &  $\rso$ \T   & $\rso$   & $\rso$ & $\rso$ &  km~s$^{-1}$     & g~cm$^{-3}$         & $\mso$             & $\mso$                            &       days          &  days     & $\mso \rm{ yr}^{-1}$ \\

\hline
20    & 10.46   &  10.20 & 0.28    & 0.08 - 0.24  &   10.74  &  $7.4\times10^{-8}$    & $3.6\times10^{-6}$     &  $5.8\times10^{-7}$       &   0.53       &  18250  &  $7.3\times10^{-8}$   \\
60    &  22.04  &   21.34 & 2.84   & 0.23 - 1.93 &   69.26  &    $6.2\times10^{-9}$  & $1.6\times10^{-5}$     &  $9.8\times10^{-7}$     &   0.61       &   1570    &  $3.7\times10^{-6}$      \\
\hline

\end{tabular}
\end{minipage}
\end{table*}

\section{Subsurface convection}\label{sub}
The occurrence and properties of the iron convection zone were studied in detail by  C09.
Here we summarize their findings and describe some of the fundamental properties of the outer regions of hot massive stars.
The evolutionary calculations used in this paper are the same as in C09. We refer to that paper and to \citet{2011A&A...530A.115B} for a detailed discussion of the models and the stellar evolution code with which these are computed.

The peak in the opacity  that causes the FeCZ occurs at around 1.5$\times 10^5$ K. 
In the envelope of a massive star this region remains relatively close to the surface during the whole main sequence. When the star expands to become a supergiant, the surface temperature decreases and the convective zone moves downwards in radius, further from the surface. 
For photospheric temperatures $\lesssim10^4$~K, partial ionization of H and He drive convection directly at the surface. 
Therefore we limit the discussion of the properties of the FeCZ  to spectral classes O and B (and partially A), which lack this additional convective layer\footnote{The discussion in Sect.~\ref{spots} is more general, and can be applied to any radiative star with a convective region below the surface.}.

The presence of subsurface convection depends also on the luminosity and metallicity of the star. At solar metallicity the FeCZ is present in models above $\sim 10^{3.2} \lso$, while at the metallicities of the LMC and SMC the thresholds are $\sim 10^{3.9} \lso$ and $\sim 10^{4.2} \lso$ respectively. Atomic diffusion and radiative acceleration can in principle lower these limits \citep{2000ApJ...529..338R}.  

Due to the very low densities in the outer layers of early type stars ($\rho \sim 10^{-8}$ g cm$^{-3}$), these convective regions contain very little mass despite their large radial extent. In Table~\ref{table} we show the properties in the outer layers of  Galactic (solar metallicity) O/B stars above the luminosity threshold. Values for the size of the convection zone and of the radiative layer above are  shown for models of a 20 and a 60 $\mso$ at ages \ttwenty yr and \tsixty yr respectively. These correspond to the calculations discussed in detail in C09, where is possible to see how  the extension and location of subsurface convection changes during the MS evolution (See their Fig.~2 and 3).

In the FeCZ  the transport of energy by convective motions is relatively inefficient: radiation dominates and transports most of the total flux (e.g. about 95\% in the 60 $\mso$ model of Table~\ref{table}). This is because the density is very low and the mean free path of photons correspondingly long. In this situation convection is significantly superadiabatic, and the actual gradient $\grad\equiv {\rm d}\ln T/{\rm d}\ln P$ has to be explicitly calculated from the mixing length equations \citep[e.g.][]{Kw90}. 
C09 found velocities ranging from  1 to 10's of $\kms$ in the FeCZ of OB stars, where the sound speed is usually of the order 100$\kms$. 

In addition, it is informative to estimate the values of the diffusivities in the FeCZ. As usual in non-degenerate stars, the thermal diffusivity is the largest: 
\begin{equation}\label{eq:chi}
\chi = \frac{K}{\rho c_{\rm p}} = \frac{4acT^3}{3\rho^2 c_{\rm p} \kappa}, 
\end{equation}
where $a$, $c_{\rm p}$ and $\kappa$ are the radiation constant, the specific heat at constant pressure and the Rosseland mean opacity. In the 20 and a 60 $\mso$ models $\chi$ is around $5\times10^{17}$ and $8\times10^{19}$ cm$^2$ s$^{-1}$ respectively. 
 The momentum diffusivity (kinematic viscosity) is (in contrast to lower mass stars) also dominated by photons, and is given in this case by
\begin{equation}
\nu = \frac{4aT^4}{15c\kappa\rho^2}
\end{equation}
where in the 20 and a 60 $\mso$ models it is approximately $6\times10^9$ and $10^{12}$ cm$^2$ s$^{-1}$. Finally, the magnetic diffusivity as given by Spitzer's expression 
\citep{1962pfig.book.....S}
\begin{equation}
\eta = \frac{\pi^{1/2}m_e^{1/2}Ze^2 c^2}{\gamma_E 8 (2k_BT)^{3/2}}\ln\Lambda \approx 7\times10^{11}\ln\Lambda \;T^{-3/2} \,{\rm cm^2}\,{\rm s}^{-1}
\end{equation}
which is roughly $10^5$ cm$^2$ s$^{-1}$ in both models (assuming a fully ionized H plasma and a value $\ln\Lambda =10$ for the Coulomb logarithm). Note that photons have a negligible effect on the magnetic diffusivity.
The magnetic Prandtl number (Pr$_{\rm m}\equiv\nu/\eta$) is consequently $10^5$ to $10^7$. We can compare these numbers to the values usually discussed in the solar dynamo (Pr$_{\rm m}\sim 10^{-2}$ in the tachocline) and   in the  galactic dynamo contexts (the ISM has typically Pr$_{\rm m}\sim 10^{12}$).


\section{Magnetic spots}\label{spots}
The occurrence of convection zones close to the surface of hot massive stars opens
a new scenario. Magnetic fields could be readily produced by dynamo action and reach the surface via magnetic buoyancy (C09). 
Below we discuss this hypothesis.
\subsection{Dynamo Action}
In an astrophysical plasma a dynamo is a configuration of the flow which is able to generate a magnetic field and sustain it against ohmic dissipation.
Depending on the scale of the resulting magnetic field respect to the scale of kinetic energy injection, dynamos are usually divided into small and large scale. 
In large scale dynamos the field has correlation length bigger than the forcing scale in the flow, while small scale dynamos result in magnetic fields with correlation scale of order or smaller than the forcing scale.
Small scale dynamos can occur in non-helical turbulent flows, while anisotropic flows (e.g. shear flows) are required for a large scale dynamo.
In the mean field approach, large scale dynamos are often divided into $\alpha\Omega$ and $\alpha^2$, depending on the role of the $\Omega$ and $\alpha$ effect in regenerating the toroidal and poloidal components of the field.  We refer to \citet{2005PhR...417....1B} for a very nice review of astrophysical dynamos.

The FeCZ is a turbulent layer close to the surface of a massive star and as such could be the site of a small scale dynamo \citep[e.g.,][]{1994IAUS..162..173M}.
Moreover main sequence massive stars usually rotate rapidly.
A typical equatorial rotational velocity of $150\kms$  \citep[typical for Galactic
B-type stars, e.g.][]{2006A&A...457..265D} corresponds to a rotational period of the
order of the convective turnover timescale (Rossby number is in the range
1...10), therefore an efficient  $\alpha^2$ or  $\alpha \Omega$-dynamo could be possible.
Assuming magnetic fields at equipartition with the kinetic energy of the
convective motion, gives magnetic fields up to  $\sim2$kG. This is supported by
simulations of turbulent convection in the presence of rotation and shear, which
show dynamo excitation with magnetic fields reaching equipartition on scales
larger than the scale of convection
\citep{2008A&A...491..353K,2010arXiv1009.4462C}. 

\begin{figure}
\resizebox{\hsize}{!}{\includegraphics{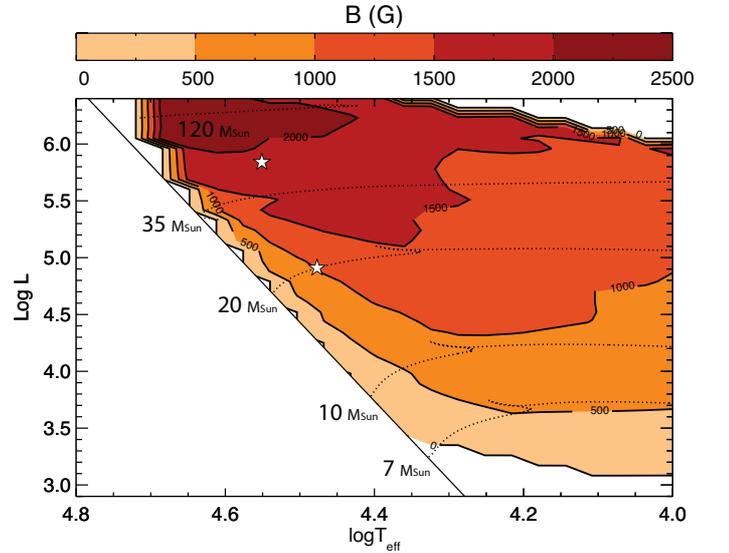}}
  \caption{Values of maximum magnetic fields (in gauss) in the FeCZ, as a function of the location in the HR diagram. This plot is based on
evolutionary models between 5$\mso$ and 120$\mso$ for  solar metallicity (some evolutionary tracks shown as dotted lines). The amplitude of magnetic fields is calculated assuming equipartition with the kinetic energy of convective motion. The full drawn black line roughly corresponds
to the zero age main sequence. The two star symbols correspond to the location of the 20 and 60$\mso$ models discussed in the text and in C09.
  }  \label{beq}
\end{figure}

Hence dynamo action in the FeCZ  could depend on parameters like the stellar
rotation (as measured by the Rossby number) and the shear profile in the region
of interest.  The scale of the magnetic field depending on the type of dynamo occurring in the relevant layers.
The dynamo may also be affected by a (presumably fossil) large-scale magnetic field penetrating upwards from the
radiative zone below (see Sect.~\ref{disc}). 
However it is interesting to note that many of the photospheric and wind phenomena observed in hot massive stars are ubiquitous 
\citep[e.g.,][]{1989ApJS...69..527H,1996A&AS..116..257K,1996ApJS..103..475F,2002A&A...388..587P}, in contrast to fossil fields which have been found at the surfaces of some (as yet badly measured) fraction of massive stars.
As it is difficult to predict the exact rotational properties of the plasma inside the FeCZ, here we will not focus on a detailed study of subsurface dynamo action. 
As a preliminary study of subsurface magnetism we will just follow the results of \citet{2009A&A...499..279C,2010arXiv1009.4462C}, and assume that magnetic fields at equipartition are produced in the FeCZ. 
In Fig.~\ref{beq} we show the expected maximum magnetic field inside the FeCZ, assuming equipartition of magnetic and kinetic energy:
\begin{equation}\label{equipartition}
\frac{B^2_{\rm eq}}{8\pi} = \frac{1}{2}\,\rho\,\vca^2,
\end{equation}
where we adopted the maximum value of the $\rho \vca^2$ 
 inside the FeCZ (as computed in the non-rotating models of C09). Due to the higher power dependency on the velocity, the location of the maximum of $\rho \vca^2$ always roughly corresponds to the maximum of $\vca$. Values of magnetic fields calculated using the average convective velocity and density typically differ by less than $30\%$. 

\subsection{Magnetic buoyancy}\label{buoyancy} 
 There are various possible mechanisms which can bring a magnetic field
generated in the FeCZ through the overlying radiative layer to the surface, and it is a useful exercise to
compare their timescales. As the title of this section suggests, it
turns out that
among the mechanisms studied here,
the dominant one is magnetic buoyancy.

The magnitude of the mass loss from massive stars is such that
material resides in the outer, radiative layer for just a short time. Assuming a
rate of $\sim 10^{-7} M_\odot$ yr$^{-1}$, this time is $\sim 1$ yr. 

Magnetic field could also be brought to the surface by Ohmic diffusion. The timescale over which it produces effects over the relevant radial length scale, i.e. the pressure scale height $\hp$, is
\begin{equation}
\tau_{Ohm} \sim \frac{\hp^2}{\eta} \sim \left(\frac{\hp}{10^{10}\,{\rm
cm}}\right)^2\,\left(\frac{T}{10^5\,{\rm K}}\right)^{3/2} \, 10^{7} \,{\rm  yr},
\end{equation}
where $\eta$ is the magnetic diffusivity. We shall now see that this timescale and that of advective transport are much longer than the buoyancy timescale.

Buoyancy causes magnetic features to rise through a radiative zone because magnetic field provides pressure without contributing to density. Since
the sound and Alfv\'en timescales are shorter than the thermal/buoyancy timescale (the buoyant rise is subsonic and sub-Alfv\'enic), we can assume that a magnetic feature is in pressure equilibrium with its surroundings. Calling the sum of gas and radiation pressure inside and outside the feature
$P_{\rm i}$ and $P_{\rm e}$ we have
\begin{equation}\label{pressure}
P_{\rm e} = P_{\rm i} + P_{\rm mag},
\end{equation}
where the value of the magnetic pressure $P_{\rm mag}$ depends on the geometry of the feature. For instance, in a self-contained structure it is $B^2/24\pi$, and in a flux tube of fixed length it is $B_{\rm ax}^2/8\pi$ where $B_{\rm ax}$ is the component along the axis of the tube\footnote{See \citet{2010MNRAS.406..705B}  for a discussion of this point and a review of the stability of flux tubes.}. Note the implicit assumption that the size of the structure $l$ is much smaller than the scale height $\hp$; it is quite likely (see below) that in fact $l\approx \hp$ but this will make a difference just of factors of order unity. Now, in the absence of thermal diffusion the feature reaches an equilibrium where $\rho_{\rm i}=\rho_{\rm e}$, made possible by the internal temperature being lower than the external $T_{\rm i}<T_{\rm e}$. With  the addition of a {\it small} thermal diffusion, the feature absorbs heat from its surroundings and rises quasistatically upwards through surroundings of
increasing entropy. The speed of this rise is given by:
\begin{equation}\label{thermal}
\varv_{\rm therm} \sim \frac{2 \chi \hp}{l^2 \beta\,(\nabla_{\rm ad}-\nabla)}\,\cdot\,\frac{1}{4-3\alpha}
\end{equation}
where $\beta\equiv P_{\rm e}/P_{\rm mag}$, $l$ is the size of the feature, 
 $\nabla$ and $\nabla_{\rm ad}$ have their usual definitions and take values $\nabla = \adrad \approx 0.24$ and $\nabla_{\rm ad} \approx 0.25$ in the radiative zones between the FeCZ and the photosphere, and $\chi$ is the thermal diffusivity given in (\ref{eq:chi}).

If on the other hand thermal diffusivity is {\it large}, the feature rises so fast that the
speed is limited by aerodynamic drag. In this regime the speed is independent of thermal conductivity, and we can make the
approximation that $T_{\rm i}=T_{\rm e}$; note that unlike above, in this regime we have $\Delta\rho\equiv\rho_{\rm e}-\rho_{\rm i}\neq0$. The buoyant force as given by
Archimedes' principle is balanced by the drag force, 
\begin{equation}
\frac{1}{2}C_{\rm d}A\rho_{\rm e} \varv_{\rm drag}^2 = Vg\Delta\rho
\end{equation}
where $A$ and $V$ are the cross-sectional area (projected onto a horizontal plane) and volume of the magnetic feature, and $C_{\rm d}$ is the drag coefficient whose value depends on geometry. This can be rearranged to
\begin{equation}
\varv_{\rm drag}^2 \approx \left(\frac{2V}{C_{\rm d}\hp A}\right) \frac{P_{\rm mag}}{\rho_{\rm e}}.
\end{equation}
Since the length scale of the feature $l\approx V/A$, the term in brackets is approximately equal to $l/\hp$.\footnote{If the feature is spherical, $C_{\rm d}\approx0.75$ \citep{2001ApJ...554..261C}.} The remaining part is approximately the square of the Alfv\'en speed, which one can alternatively express in terms of the sound speed $c_{\rm s}$ and the ratio of total to magnetic pressure, $\beta$:
\begin{equation}\label{drag}
\varv_{\rm drag}\sim \varv_{\rm A} \left(\frac{l}{\hp}\right)^{1/2} \sim \frac{c_{\rm s}}{\beta^{1/2}}\left(\frac{l}{\hp}\right)^{1/2}.
\end{equation}
Putting the numbers into (\ref{thermal}) and
(\ref{drag}) gives
\begin{eqnarray}
\varv_{\rm therm} &\approx& \left(\frac{l}{\hp}\right)^{-2} \beta^{-1}\;\, 10^{10}\,-\,10^{13} \;{\rm cm\,s}^{-1},\\
\varv_{\rm drag} &\approx& \left(\frac{l}{\hp}\right)^{1/2} \beta^{-1/2}\;\, 10^{7}\,-\,3\times10^{7}
\;{\rm cm\,s}^{-1},
\end{eqnarray}
where ranges in the numbers come from the difference between 20 and 60 $\mso$ models and from differences between top and bottom of the surface radiative layer. It is clear that unless $l\gg \hp$, for which the approach above breaks down anyway, or the field is implausibly weak (and therefore undetectable and unimportant), $\varv_{\rm therm}$ is larger and the rise is therefore limited by drag. For a magnetic feature with $l\sim \hp$ and $\beta\sim100$ (a conservative estimate), the time taken to rise one scale height is of order 2 hours in both models, much shorter than the advective and Ohmic timescales.

This is in contrast to the thermal-diffusion-limited regime
which was studied by \citet{2003ApJ...586..480M} 
to check the possibility that magnetic fields generated in a convective core could rise to the surface. In that
case the timescale appears to exceed the main sequence lifetime unless
very small-scale features are used. Moreover \citet{2005MNRAS.356.1139M} 
argue that the inclusion of compositional gradients may suppress the buoyancy,
making this scenario unlikely\footnote{\citet{2005MNRAS.356.1139M} also discuss
the possibility of buoyant rise of magnetic fields generated in the
radiative zone by the Tayler-Spruit dynamo. However it is not clear how this
dynamo process actually works 
\citep{2006A&A...449..451B,2007A&A...474..145Z}.}.
    
The situation in massive star surface layers is in some
sense more similar to that of flux tubes in the solar convection zone
in that the rise is limited by drag forces, although the drag force in
the solar CZ does not scale simply as $v^2$ but depends also on the
convective motion. 
A stationary flux tube in the solar CZ experiences a
net downwards aerodynamic force once the average of upwards- and downwards-moving
convective cells has been taken, with the result that there is a
field-strength threshold below which the tube moves downwards despite
its intrinsic buoyancy. 
 
 Once a magnetic field element reaches the surface of the star 
it would be useful to estimate its lifetime \tauspot.
This is not trivial, as the photosphere of OB stars is complicated by  the presence
of strong winds and, likely, turbulence and shear. However we can put some limits on the lifetime.

As we discussed at the beginning of this section, if the material in the radiative layer above the FeCZ has a mass $\Delta M$, it will be removed by mass loss in a time $\Delta M/\dot{M}$. 
This sets an upper limit for the time a magnetic element, which is not anchored to the convective zone, can exist at the surface: this time is  about 8 years for the 20$\mso$ and 3 months  for the 60$\mso$ model.
Of course this holds true only as long as the field is not annihilated by dissipative processes like magnetic reconnection.
If the magnetic element is anchored to the convective zone, it can exist  at the surface as long as the stellar wind does not remove all the mass contained in both the radiative layer {\em and} the FeCZ. It turns out that this limits $t_{\mathrm{spot}}$  to be less than 50 years (20$\mso$ model) and 4 years (60$\mso$ model). We refer to Table~1 in C09 for the data used for these simple estimates.
These are solid upper limits, and it is likely that such features would survive for a much shorter time, owing to turbulence, shear and magnetic reconnection.
As a lower limit for $t_{\mathrm{spot}}$ we could assume the time a feature of size $l$ takes to cross the photosphere while rising with a velocity $v_{\rm drag}$. For $l \sim \hp$ this gives a timescale of the order of hours. 

 Finally, note that it is not inconceivable that the magnetic dynamo does not saturate properly because of the finite time the gas spends in the FeCZ. However, we consider this unlikely, since the convective turnover times ($\sim$ hours - a day) appear to be very much shorter than the aforementioned mass-flux timescales of 50 and 4 years.

\begin{figure}
\resizebox{\hsize}{!}{\includegraphics{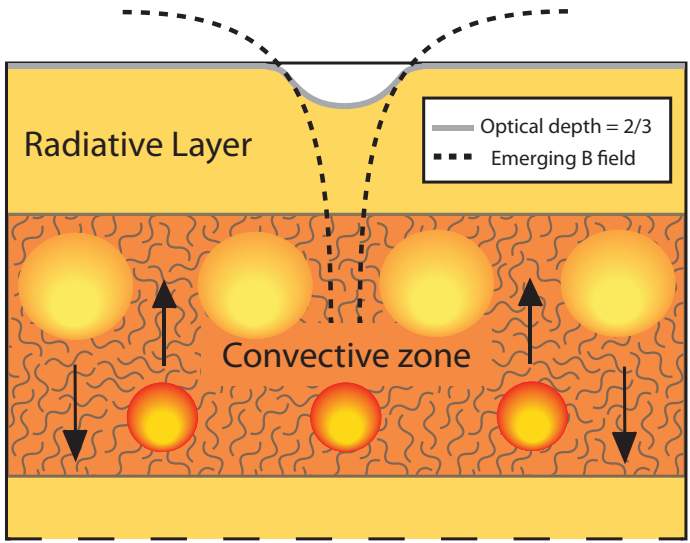}}
  \caption{Schematic representation of the effect of an emerging magnetic element at the surface of a hot massive star. 
  A magnetic field (dashed line) rising from the subsurface convection zone threads the radiative layer and reaches the stellar photosphere.  
  The solid, grey line shows the location of the stellar surface, as defined by the value 2/3 for the optical depth. 
  Notice that inside the magnetic field this line reaches deeper (and hotter) layers of the star, as compared to regions not affected by the field. }
  \label{sketch}
\end{figure}

\section{Observable effects}\label{consequences}
Let us first estimate the likely field strength at the photosphere. Whilst one might expect it to be proportional to the field strength produced in the convective layer, the difference in field strength of a magnetic feature between the top and bottom of the radiative layer depends on its geometry. In a self-contained magnetic feature (a `blob' or `plasmoid') the field strength scales as $B\propto \rho^{2/3}$, while a horizontal flux tube of fixed length has $B_{\rm ax}\propto\rho$ and $B_{\rm h}\propto \rho^{1/2}$ where $B_{\rm ax}$ and $B_{\rm h}$ are the axial and `hoop' components. Knowing that $P\sim\rho^{4/3}$ (approximately, since $\nabla\approx1/4$) in the radiative layer, we see that these two scenarios give $\beta=\,$const and $\beta\propto P^{-1/2}$ respectively. In constrast, if the central section of a flux tube rises to the surface while its ends are still in the convective layer, plasma can flow downwards along the tube and the field strength at the surface can in principle reach equipartition with the surrounding 
(gas plus radiation) pressure, as we see happening in the solar CZ.
Assuming equipartition magnetic field in the convective layer gives up to
$\approx 2$ kG at the base of the radiative layer, and $\beta\approx 100$. The
density contrast across the radiative layer is around $50$ and $100$ in the $20
M_\odot$ and $60 M_\odot$ models respectively, which gives field strengths at
the photosphere of $40-150$ G and $20-100$ G respectively in the blob and
horizontal tube cases. The photospheric field could be significantly greater if a tube arches upwards and fluid is allowed to flow along it --  the limit should then be of the order of equipartition with the photospheric pressure (about $300$ G) or somewhat more, since the photosphere (i.e. the $\tau=2/3$ level) in the tube is lower.

\begin{figure}
\resizebox{\hsize}{!}{\includegraphics{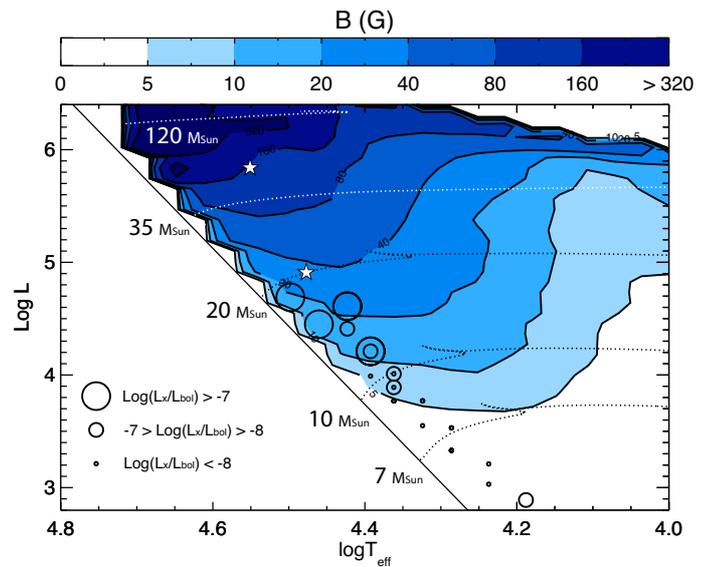}}
  \caption{Minimum values of expected surface magnetic fields (in gauss) 
as a function of location in the HR diagram. This plot is based on
evolutionary models between 5$\mso$ and 120$\mso$ for Galactic metallicity (some shown as dotted lines). Surface magnetic fields are calculated scaling 
the equipartition fields in the FeCZ (see Fig.~\ref{beq})  with $\rho^{2/3}$ (see discussion in Sect.~\ref{consequences}). The full drawn black line roughly corresponds
to the zero age main sequence. Circles  correspond to the observed stars in the  \citet{1997ApJ...487..867C} sample, with symbol size coding different range of $L_{\rm x}/L_{\rm bol}$, as explained in the legend. The two star symbols correspond to the location of the 20 and 60$\mso$ models discussed in the text and in C09. }  \label{bsur}
\end{figure}

 In Fig.~\ref{bsur} we show the expected amplitude of surface magnetic fields
emerging from the FeCZ, assuming a scaling $B\propto \rho^{2/3}$. 
We argue
that this represents a lower limit for the magnetic field that should be found
at the surface of hot massive stars, if equipartition of kinetic and magnetic
energy is reached in their FeCZ. 
It might seem plausible that magnetic features rising through the radiative layer partially annihilate each other on the way up. However, if these magnetic bubbles have a size at the bottom of the radiative layer comparable to the local scale height, they only have to rise through a distance about three times their own size to reach the photosphere, expanding as they do so. They rise at approximately the Alfv\'en speed, and since any reconnection would presumably take place at some fraction of the Alfv\'en speed, there is limited time available to destroy much of the original flux. It seems more likely that significant reconnection does take place {\it above} the photosphere.

While these fields appear to have very small amplitude, their role at the
stellar surface should not be underestimated. In Fig.~\ref{bsur},
the ratio of the magnetic to wind pressure \citep[$\eta\approx B_{\rm eq}^2
R_*^2 / \dot M \vin^2$,][]{2002ApJ...576..413U}
is of the order of unity over
the whole region where magnetic fields are present. 
This means that already such low magnetic fields can alter the dynamic of
the stellar wind, for example contributing to seed instabilities in the flow.

\subsection{Surface bright spots and photometric variability}\label{photometric}
Once a magnetic field element reaches the stellar photosphere, one would like to
know what effect this can have on observable properties of the star. In the Sun,
magnetic fields of amplitude up to $\sim3$ kG emerge at the surface as sunspots.
 The main effect of such magnetic fields is to locally inhibit convection. As
the energy flux below the surface of the Sun is mainly transported by
convection, this produces a decrease of the flux at the surface, resulting in a
visible dark spot. In contrast, the hot massive stars considered here have
radiative surfaces and the effect of an emerging magnetic field  is very
different.

Under the simple assumption of magnetohydrostatic equilibrium we can make use of  Eq.~\ref{pressure}. Assuming again that 
also the gas temperature inside and outside is the same ($T_e = T_i$),  we
obtain once more that the density inside the magnetic element is lower than
outside. Since we are at the stellar surface, this makes the element more
``transparent'', i.e. photons can escape from deeper down the star compared to
outside the magnetic flux.
This is somehow similar to what occurs in the solar faculae,
which are regions of emerging magnetic fields with a scale much smaller than
the scale of convection. In that case \citet{2004ApJ...607L..59K}  found, using MHD
calculations, that ``The opacity in the magnetic flux concentration is strongly
reduced owing to its low density and temperature and thus provides a clear
sight straight through the magnetic field onto the adjacent nonmagnetic
granule''. Their calculations basically verified the explanation for solar
faculae proposed by \citet{1976SoPh...50..269S,1977PhDT.......237S}.

Photons emerging from deeper down the star are emitted from regions where
the temperature is higher. 
The temperature stratification, assuming that the magnetic pressure is much
smaller than the  
gas $+$ radiation pressure (i.e. $\beta \gg 1$), 
is given by the radiative
gradient $\adrad$
\begin{displaymath}
\adrad \equiv \Bigg( \frac{{\rm d}\ln{T}}{{\rm d}\ln{P}}\Bigg)_{\mathrm{rad}} .
\end{displaymath}
Therefore it is easy to estimate the temperature difference between the region of the star visible through the magnetic element and outside of it:
\begin{displaymath}
\frac{\Delta T}{\Delta P} = \frac{T\Delta(\ln{T})}{P \Delta(\ln{P})} \approx \frac{T}{P} \adrad \,
\end{displaymath}
\begin{equation}\label{temp}
\frac{\Delta T}{T}\approx\frac{\adrad}{\beta} \, .
\end{equation}
This means that regions with an emerging magnetic field will look hotter.  
In the case of the FeCZ, assuming the minimum value of magnetic fields shown in Fig.~\ref{bsur}, preliminary estimates give temperature differences up to a few hundred K. 
In the case of a surface magnetic field in equipartition with the photospheric pressure (gas plus radiation) equation~\ref{temp} 
is not strictly correct, as the magnetic field can affect the radiative gradient. 
Nevertheless, assuming for a moment that the correction 
to this relation is small, one expects temperature differences up to a few kK.  

Since the stellar luminosity is related to the effective temperature through the
relation $L=4\pi  R^2 \sigma T^4$, magnetic fields in stars having radiative
surfaces will produce bright spots. We can estimate the local luminosity
contrast between the magnetic bright spot and the non magnetic regions of the
stellar surface
\begin{equation}
\frac{\Delta L_{\rm loc}}{L_{\rm loc}}\approx4\, \bigg(\frac{\Delta T}{T}\bigg) \approx  \frac{4\,\adrad}{\beta} \, .
\end{equation}
Temporal changes in the observed stellar luminosity might be expected, as a result of the appearance/disappearance 
of such a bright spot. 
In a rotating star the luminosity would also decrease (increase) when the spot moves out of (into) the visible part of the stellar disc. It is difficult at this stage to give a firm estimate of the total expected change in luminosity, since this will depend not only on the intensity of the magnetic field, but also on the geometry and number of magnetic spots. 
Also, it is not clear exactly where the extra luminosity in the spots should come from, in terms of underluminous areas. 
Nevertheless, if we assume a circular spot of radius $r$, with filling factor $f =  (r/R)^2$, where $R$ is the stellar radius, then the effect on the stellar luminosity can be estimated if $r$ is known. A simple approach, based on preliminary models of  subsurface convection \citep{2010arXiv1009.4462C}, is to equate $r$ to the pressure scale height in the FeCZ. These simulations show dynamo action producing magnetic fields on scales comparable to or larger than the scale of convection, i.e. the local pressure scale height $\hp$.
We obtain:
  \begin{equation}
\frac{\Delta L}{L}=  \frac{4\, \adrad}{\beta} \, \bigg( \frac{\hp}{R}\bigg)^2\,
,
\end{equation}
which, for typical values of subsurface convection, gives $\Delta L /L \sim
10^{-5}$. This is the relative change in luminosity caused by the
appearance/disappearance of one magnetic spot of about 100 G with
size 0.2$\rso$ at the surface of a hot massive star. 
Recall that these stars have radii around $10R_\odot$.

Of course spots with stronger magnetic fields and bigger filling
factors are certainly possible. Such spots would lead, most likely, to larger
changes in the integrated light from the star. An estimate of the
amplitude of the variability as function of filling factor and surface
strength is much more difficult in this case, as one can no
longer consider the magnetic element thin (in a thermal sense). MHD calculations
of magnetic flux tubes in a radiative environment are required to study the
details of this problem. 

Photometric variability at the micro level ($10^{-3}$ magnitudes) and with
timescales of the order of days seems to be widespread in O stars \citep{1992MNRAS.254..404B}. 
Using COROT to perform high-precision photometry, \citet{2010A&A...519A..38D}  found
solar-like oscillations in a young O-type star (HD46149). 
These are modes with a finite lifetime that are believed to be excited by the
presence of subsurface convection \citep{2010A&A...510A...6B}.
Intriguingly, this system also shows a low-frequency variability with time scale of days and amplitude of order $10^{-3}$ magnitudes.
The authors could not relate this variability to any type of pulsations. Instead they argue that this could be the signature of spots, stellar winds or chemical inhomogeneities 
modulated by the stellar rotation \citep{2010A&A...519A..38D}.
\citet{2011A&A...528A.123P} obtained high-precision photometric observations of the B0.5IV
star HD51756 with COROT. While no solar-like oscillations were identified in
this case, they found cusp-like features in the light-curve. 
These features have amplitudes of order  $10^{-3}$ magnitudes, and recur on a
time scale of days.  Similar to HD46149, \citet{2011A&A...528A.123P} propose as
a possible source for this variability photospheric features like
spots or chemical inhomogeneities, modulated by stellar rotation. 

\subsection{X rays} 
OB stars have been found to emit X rays \citep{1979ApJ...234L..51H}. O stars  follow the relation $L_{\rm x}/L_{\rm bol}  \sim 10^{-7}$ \citep[see e.g.][]{1981ApJ...248..279P} while B star X-ray luminosities show a lot of scatter
\citep{1992A&A...265L..41M,1997A&A...322..167B}. Among possible physical mechanisms for the production of X rays are magnetic coronae \citep{1979ApJ...229..304C,2007ApJ...668..456W,2009ApJ...692L..76W}  and the intrinsic instability of line-driven winds \citep{1980ApJ...241..300L,1988ApJ...335..914O,1995A&A...299..523F}. 
 
The $L_{\rm x}/L_{\rm bol}$ relation only applies for $L_{\rm bol} > 10^{38}$ erg s$^{-1}$, i.e. earlier than B0-B1 type stars. It breaks down at lower luminosities, with  the ratio $L_{\rm x}/L_{\rm bol}$ two orders of magnitude smaller at B3 than at B1 \citep{1994ApJ...421..705C,1997ApJ...487..867C}. This is shown in Fig.~\ref{bsur}, where we plot the data from  \citet{1997ApJ...487..867C}  together with the theoretical expectation for the minimum values of surface magnetic fields (assuming dynamo action in the FeCZ). While there could be different explanations for such a correlation, it is interesting to note that in the sample stars with  $L_{\rm x}/L_{\rm bol} > -7$ correspond to regions of the HRD with larger values of predicted surface magnetic field. 
Recently  \citet{2011ApJS..194....5G} have studied with {\it Chandra} the X-ray emission from OB stars in the Carina Complex. They also found that most of their O stars follow the  $L_{\rm x}/L_{\rm bol}$ relation, while  this  breaks down for B stars. Interestingly they identify a group of B stars with high X-ray emission that cannot be explained by a distribution of ordinary coronal, pre main-sequence companions. A possibility is that in O stars wind embedded shocks dominate and are responsible for the $L_{\rm x}/L_{\rm bol}$ relation,  while the observed X-ray luminosity in B stars is mainly due to the presence of subsurface generated magnetic fields.


\section{Discussion}\label{disc}
We showed that, among advection, magnetic diffusion and magnetic buoyancy, the
most likely process that can bring to surface magnetic fields generated by dynamo action in the FeCZ is magnetic buoyancy.
However there could be other ways for a magnetic field to escape the subsurface convection region and reach the stellar surface.
For example, \citet{2010A&A...523A..19W,2011arXiv1104.0664W} studied the magnetic flux produced by a turbulent dynamo in Cartesian and spherical geometry, respectively.
They found that magnetic flux can rise above the turbulent region without the need of magnetic buoyancy. 
This appears to be related to the release of magnetic tension, which leads to the relaxation and emergence of the field.
Convective overshoot might also contribute to the transport of magnetic flux out of the FeCZ.
However, given the fact that the FeCZ usually sits a few pressure scale heights from the stellar surface, this effect is likely marginal.


Some fraction of massive stars display large-scale, apparently fossil fields of strengths ($1-3$ kG) comparable to the FeCZ equipartition strengths derived above. A sufficiently strong field will suppress convection, forcing an increase in temperature gradient so that the entire energy flux can be transported radiatively. However, it is not obvious where the field strength threshold should be; one might na\"{i}vely expect convection to be suppressed by a field of greater energy density than the convective motion, but the work of \citet{1966MNRAS.133...85G,1969MNRAS.145..217M,1970MSRSL..19..167M}
suggests that in that case the temperature gradient would simply steepen further above the adiabatic gradient until convection resumes, and that to suppress convection completely, a field above (approximate) equipartition with the {\it thermal} energy would be required. However, this work considers a situation where the energy flux is almost entirely convective, such as in the bulk of the solar convective zone. In the FeCZ context the situation is different, in that only a small fraction  of the stellar heat flux is carried by convection, and that the temperature gradient is already significantly above adiabatic. It is plausible therefore that a fossil field of a few kG 
could indeed suppress convection, which would have consequences on observables connected to the presence of subsurface convection (e.g. microturbulence, C09).  One can make an analogy here with the intermediate-mass stars (A and late B), where fossil fields do appear to suppress or at least significantly reduce the weak helium-ionisation convection at the photosphere which manifests itself as microturbulence (D.\,Shulyak, priv.\,comm.)

C09 proposed the FeCZ as a possible unifying mechanism for the formation (or seed) of different observational phenomena at the surface of hot, massive stars.
The presence of turbulent subsurface convection appears to be able to trigger at least three different classes of physical phenomena: running waves \citep[g-modes and p-modes, e.g.][C09]{1990ApJ...363..694G},
solar-like oscillations \citep[e.g., C09;][]{2010A&A...510A...6B} and magnetic fields (C09; this work). Each of these perturbations can, in principle, results in observable effects at the surface and/or in the stellar wind.
Estimating the amplitude, length-scale and  time-scale of the surface perturbations produced by each mechanism is of primary importance to establish the connection between FeCZ and observable phenomena.
For example wind clumping could be produced by small-scale velocity and/or  density fluctuations at the stellar surface. These perturbations could excite the line-driven instability \citep{2002A&A...381.1015R} closer to the stellar surface than originally predicted from the theory. This would reconcile theory with observations of the radial stratification of wind clumping \citep{2006A&A...454..625P}. However both running waves and localized magnetic fields could be responsible for such a perturbation at the base of the wind. To understand which one dominates, hydrodynamic simulations of the line-driven instability are required. These calculations need to include, as boundary conditions at the base of the stellar wind, the different velocity and density perturbations predicted in the case of  running waves and surface magnetic fields. While simple estimates for these perturbations exist (C09 and this paper), realistic MHD simulations of the FeCZ and the radiative layer above it are necessary. 
Understanding the exact role and properties of each perturbation is also fundamental to explain the puzzling coexistence of small- and large- scale phenomena observed at the surface and in the wind of hot, massive stars (e.g. microturbulence/macroturbulence and wind clumping/DACs). An effort is clearly required to improve existing MHD calculations of subsurface convection \citep{2010arXiv1009.4462C}.

The presence of surface magnetic fields  might have implications for the evolution of OB stars. 
This is because, for wind confinement parameters $\eta \ga 1$, the magnetic field  can in principle affect the stellar wind mass-loss and angular momentum-loss rates.
For large scale fields at the surface, the impact on angular-momentum loss has been studied \citep{1967ApJ...148..217W,2009MNRAS.392.1022U}. In the case of   the magnetic star $\sigma$~Ori~E the theoretical spin-down is in agreement with observations \citep{2010ApJ...714L.318T}.  The effect of large scale fields on the angular momentum evolution of main sequence massive stars has been investigated by 
\citet{2011A&A...525L..11M}. Their results seem to show that only fields above a few hundred gauss can have a substantial impact. For a given surface field amplitude, higher order multipoles (i.e. less coherent fields, as expected in the case of dynamo action in the FeCZ)  result in a reduced angular momentum-loss. 
Moreover in the case of small-scale fields a lot of the flux may reconnect and disappear fairly close to the surface, below where most of the wind acceleration takes place. Therefore we expect  magnetic fields produced in the  FeCZ to have only a modest effect on the angular momentum evolution of massive stars.  Nevertheless further investigation of their geometry and amplitude is required to confirm this statement.

\section{Conclusions}\label{conc}
We have shown for the first time that emerging magnetic fields produced
in the FeCZ of hot, massive stars could reach the stellar surface. The surface fields expected are localized and with amplitudes smaller than a few
hundred gauss. 

As pointed out by \citet{2005ASPC..337..114H}, a number of unexplained observational
phenomena in hot, massive stars are consistent with surface magnetism.
These phenomena include the photometric variability and X-ray emission discussed above, line
profile variability \citep{1996ApJS..103..475F,2004MNRAS.351..552M} and discrete
absorption components \citep[DACs, e.g.][]{1996ApJ...462..469C,1997A&A...327..281K}.
The  magnetic fields emerging from the FeCZ could be widespread in OB stars at solar metallicity,
and play an important role in some or even all these phenomena. The metallicity dependency of the FeCZ is such that these effects
would weaken and eventually disappear with decreasing metallicity (C09).


Direct detection of magnetic fields emerging from the FeCZ has to overcome 
the problem that, for complex fields, opposite line-of-sight magnetic polarities and their respective opposite circular polarization signals 
do cancel when integrated over the stellar disc. However if the star is rotating fast enough,
it might still be possible to detect such fields using the Zeeman Doppler Imaging technique \citep[ZDI,][]{1989A&A...225..456S}.
While direct measurement of FeCZ-generated magnetic fields is challenging, we argue that a viable indirect approach 
is to use high-precision photometry -- indeed we show that magnetic fields at the stellar surface of radiative stars should 
produce bright spots. The finite lifetime of the magnetic spots, together with rotational modulation, could leave a detectable signature in the light curve of hot, massive stars.
This kind of  variability seems to be present in some COROT targets \citep{2010A&A...519A..38D,2011A&A...528A.123P}, and might be used to gather preliminary informations on the magnetic field geometry. 
These information could be used to assess the feasibility and the requirements of a targeted direct detection campaign.


\begin{acknowledgements} We are grateful to Norbert Langer, David Moss, Huib Henrichs, Greg Wade,  Stan Owocki, David Cohen, Marc Gagn\'e, Hilding Neilson, Enrico Moreno M\'endez, Axel Brandenburg, Petri K\"apyl\"a and Fabio Del Sordo for useful discussions and suggestions.
\end{acknowledgements}

\bibliographystyle{aa} 
\bibliography{ref} 

\end{document}